\documentclass[usenatbib]{mn2e}
\usepackage[papersize={8.5in,11in}]{geometry}
\usepackage{epsfig,graphicx}
\usepackage{amsmath,amssymb}
\usepackage{enumitem}

\newcommand{\be}{\begin{equation}}
\newcommand{\ee}{\end{equation}}

\begin{document}
\title{Supermassive Black Hole Accretion and Growth}
\author[J. W. Moffat]{J. W. Moffat\\
Perimeter Institute for Theoretical Physics, Waterloo, Ontario N2L 2Y5, Canada}
\maketitle


\begin{abstract}
The formation, accretion and growth of supermassive black holes in the early universe are investigated. The accretion rate ${\dot M}$ is calculated using the Bondi accretion rate onto black holes. Starting with initial seed black holes with masses $M_{\rm BH}\sim 10^2-10^3M_{\odot}$, the Bondi accretion rate can evolve into a supermassive black hole with masses $M_{\rm BH}\sim 10^9-10^{10}M_{\odot}$ and with a young quasar lifetime $\sim 10^5-10^6$ years by super-Eddington accretion.
\end{abstract}

\maketitle


\section{Introduction}

A central problem in the evolution of black holes is the origin of the mass gap between intermediary black hole masses $M_{\rm BH}\sim 50 -- 100M_{\odot}$ and the supermassive black holes (SMBH) with mass $M_{\rm BH}\sim 10^{10}M_{\odot}$. The formation and growth of SMBHs has been a challenging problem. The problem has been further complicated by the discovery of high redshift quasars hosting SMBHs with masses exceeding $M_{BH}\sim 10^9M_{\odot}$ as early as $\lesssim 1$ Gyr after the Big Bang
~\citep{Mortlock2011,Venemans2018,Onoue2019,Eilers2020}. The problem is exacerbated further by the discovery of high red shift quasars, $z\sim 6 -7$ as young as of order $10^5$ years~\citep{Eilers2020}. How SMBHs can grown and form in such a short time remains a mystery.

Assuming Eddington limited accretion rates and a constant source of material SMBHs in the lifetime $t_Q$ of quasars grow exponentially:
\be
M_{BH}(t_Q)\sim M_{\rm seed}\exp\biggl(\frac{t_Q}{t_S}\biggr),
\ee
where the initial seed mass $M_{\rm seed}$ denotes the mass of the black hole before the beginning of quasar activity and black hole growth, and $t_Q$ denotes the lifetime of the quasar. The Salpeter or e-folding time $t_S$~\citep{Salpeter1964}, denotes the typical time scale of the black hole growth:
\be
t_S\sim 4.5\times 10^7\biggl(\frac{f}{0.1}\biggr)\biggl(\frac{L_{\rm bol}}{L_{\rm Edd}}\biggr)^{-1}\, {\rm yr},
\ee
where $f$ denotes the accretion radiative efficiency of order $10\%$ in thin disk models~\citep{ShakuraSunyaev1973} and $L_{\rm bol}$ denotes the bolometric luminosity. According to a theoretical model based on disk accretion, it takes about 16-efoldings, or the age of the universe to grow a SMBH with a mass $10^9M_{\odot}$ from an initial seed mass $M\sim 100M_{\odot}$, with a continuous accretion rate at the Eddington limit~\citep{Volonteri2010}. This growth time depends on whether quasars follow this exponential growth rate and the assumed disk accretion mechanism. In the following, we adopt the spherically symmetric Bondi accretion infall on the quasar embedded black hole~\citep{Bondi1952,BondiHoyle1944,Weinberg2020} that accretes matter from the ambient environment in the quasar.

\section{Bondi Accretion}

For a black hole the accretion disk arises when the angular momentum as well as mass is accreted by the black hole from the surrounding medium. The inner radius $R_{\rm accr}$ is the minimum radius of a stable circular photon orbit around the spherically symmetric black hole. In General relativity (GR) for a static Schwarzschild black hole, it is the photosphere radius $R_{\rm photo}=6GM/c^2$. An outer accreting radius is the Bondi radius, $R_B\sim GM/c_s^2$, where $c_s$ is the speed of sound. We shall assume that the initial seed black hole has an accretion disk with a small angular momentum and the environmental mass surrounding the black hole is at rest carrying little angular momentum with its mass. In addition to accretion from a thin disk, there will be accretion onto the black hole from the steady spherically symmetric infall of ambient gas and matter from a large distance from the black hole at rest, with position independent pressure and density. In contrast to the attractive gravitation being resisted by angular momentum, it is now resisted by gas pressure, and we can neglect viscosity in the absence of differential rotation. This spherically symmetric accretion infall was first treated by Bondi~\citep{Bondi1952,BondiHoyle1944,Weinberg2020}.

We assume that the environmental gas and matter medium is sufficiently far from the black hole horizon radius and the radius of the inner stable photosphere orbit that we can use the Navier-Stokes equation for a non-relativistic fluid in the absence of viscosity:
\be
\frac{\partial v_i}{\partial t}+v_j\frac{\partial v_i}{\partial x_j}=-\frac{\partial\phi}{\partial x_i}-\frac{1}{\rho}\frac{\partial p}{\partial x_i},
\ee
where $v$ is the fluid velocity, $\rho$ and $p$ are the density and pressure and $\phi$ is the gravitational potential. We treat only radial flow for steady state spherically symmetric flow when the velocity has only a radial dependence $v=v_r$. The equation of continuity can be expressed as
\be
\label{cont}
v\frac{dv}{dr}=-\frac{d\phi}{dr}-\frac{1}{\rho}\frac{dp}{dr}.
\ee
We can now obtain the equation for the accretion rate:
\be
\dot M=-4\pi r^2\rho v,
\ee
where $\dot M=dM/dt$ is the r dependent accretion rate and $v$ is negative for accretion.

For the environmental gas, we assume the polytropic equation of state:
\be
p=K\rho^\Gamma,
\ee
where $K$ and $\Gamma$ are constants and $\Gamma$ is the polytropic index. We can obtain from Eq.(\ref{cont}):
\be
\frac{d}{dr}\biggl(\frac{v^2(r)}{2}+\phi(r)+\frac{K\Gamma\rho^{\Gamma-1}(r)}{\Gamma-1}\biggr)=0,
\ee
where we adopt the Newtonian potential $\phi=-GM/r$. At an infinite distance from the black hole the density and the speed of sound $c_s(r)$ approach constant values, $\rho(\infty)$ and $c_s(\infty)$, and $\phi(r)$ and $v(r)$ approach zero as $r\rightarrow\infty$.

For a polytrope the ambient speed of sound is given by
\be
c_s(r)=\biggl(\frac{\partial p}{\partial\rho}\biggr)^{1/2}=c_s(\infty)\biggl(\frac{\rho(r)}{\rho(\infty)}\biggr)^{(\Gamma-1)/2}.
\ee
To avoid transonic gas speed $v(r) > c_s(r)$ as $r\rightarrow 0$, there must be a critical radius $r_c$ for which $v(r_c)=-c_s(r_c)$. The accretion rate transonic speed is stable at a critical radius $r=r_c$ satisfying~\citep{Weinberg2020}:
\be
r_c=\frac{GM}{2c_s^2(r_c)}.
\ee
The accretion rate becomes
\begin{align*}
\dot M&=4\pi r_c^2\rho(r_c)c_s(r_c)\\&=4\pi\biggl(\frac{GM}{2r_c^2(r_c)}\biggr)^2\rho(\infty)\biggl(\frac{c_s(r_c)}{c_s(\infty)}\biggr)^{2/(\Gamma-1)}c_s(r_c)\\
& =\frac{\pi G^2M^2\rho(\infty)}{c_s^3(\infty)}\biggl(\frac{c_s(r_c)}{c_s(\infty)}\biggr)^{(5-3\Gamma)/(\Gamma-1)}.\
\end{align*}
Using the equation:
\be
\label{csEq}
c_s^2(r_c)=\biggl(\frac{2}{5-3\Gamma}\biggr)c_s^2(\infty),
\ee
we get
\be
\label{AccrRate}
\dot M=\frac{\pi\beta G^2M^2\rho(\infty)}{c_s^3(\infty)},
\ee
where
\be
\beta=\biggl(\frac{2}{5-3\Gamma}\biggr)^{(5-3\Gamma)/2(\Gamma-1)}.
\ee
The stability of a polytrop gas requires that $\Gamma\geq 4/3$. Choosing $\Gamma=4/3$, Eq. (\ref{csEq}) becomes
\be
c_s^2(r_c)=2c_s^2(\infty),
\ee
and $\beta=2.83$.

\section{Supermassive Black Hole Growth and Evolution}

We expect that in the early universe when the first galaxies and stars and quasars emerge that the proto-galaxies and quasars are sufficiently dense to create the necessary conditions for gas and material clumping to occur, producing the ambient density environment conducive to rapid stellar and black hole growth. For a SMBH to grow to a black hole with a mass $10^9-10^{10}M_{\odot}$ in the short time after the Big Bang, and to account for the short lifetime of detected quasars, the reservoir of ambient accreting gas and matter must be available over the time period of growth. A rough estimate of this required reservoir of gas and material can be made for a spherical homogeneous cloud by the expression:
\be
\label{density}
\rho\sim \frac{3M}{4\pi R^3},
\ee
where $R$ is the radius of the proto-cloud. Measurements of quasar sizes have been carried out using microlensing techniques~\citep{Matthew2020,Rojas2020} and measuring the light time travel from one side of a quasar to the other. Because of the small size of quasars, it is difficult to get accurate size measurements and determine structural information. The radial size of a quasar is expected to be $R\sim 1$ light-month -- 1 light-year. For $R\sim 1$ light year and for a SMBH with mass $M_{\rm BH}=10^9M_{\odot}$, we obtain
\be
\rho\sim 5.61\times 10^{-13}\,{\rm g/cm^3}.
\ee

The initial seed masses of black hole with $M_{\rm BH}\sim 10^2-10^3M_{\odot}$ can have their origin in remnants of massive Population III stars~\citep{Bromm2013,Valiante2016}. Another possible origin of seeds with $M_{\rm BH}\sim 10^3-10^4$ is from the direct collapse of supermassive stars in proto-galaxies ~\citep{BrommLoeb2003,Pacucci2015,Ferrara2014,Valiante2016}. Stellar-dynamical processes can allow black holes to form in nuclear clusters of second-generation stars with masses $M\sim 10^2-10^3M_{\rm \odot}$~\citep{Devecchi2012}. Primordial black holes with masses ranging up to $10^3 - 10^5M_{\odot}$ could have formed in the early universe well before galaxy formation~\citep{Khlopov2005}.

For spherical infall of gas on a black hole, we have to account for the Eddington limit on luminosity. For accretion on a mass $M$, if the gravitational attractive force does not exceed the radiation gas pressure, the luminosity $L$ cannot exceed $L_{\rm Edd}=4\pi cGM/\kappa_T$, where $\kappa_T$ is the opacity due to the Thompson scattering cross section times the number of free electrons per gram. For completely ionized hydrogen this is
\be
\label{Eddlimit}
L_{\rm Edd}=3.25\times 10^4L_{\odot}(M/M_{\odot})=1.25\times 10^{38}\biggl(\frac{M}{M_{\odot}}\biggr)\, {\rm erg/s}.
\ee

The Bondi accretion radius for the seed black holes in quasars, $R_{\rm accr}$, is small compared to the sizes of the early universe quasars $R\sim 0.5-1$ pc. We shall choose for the ambient gas and material density of a quasar $\rho(\infty)=10^{-10}-10^{-12}\,{\rm g/cm^3}$. We choose for the density $\rho(\infty)=5\times 10^{-11}\, {\rm g/cm^3}$, the speed of sound $c_s(\infty)=10\,{\rm km/s}$, an initial seed mass $M_{\rm BH}=10^3M_{\odot}$ and $\beta=2.83$.  We obtain from (\ref{AccrRate}):
\be
\label{accreetgpersec}
\dot M=7.27\times 10^{30}\, {\rm g/s},
\ee
and per year it is
\be
\label{accret1}
\dot M=3.67\times 10^3M_{\rm \odot}/{\rm yr}.
\ee
For a quasar lifetime $t_Q\sim 10^6$ yr, we get
\be
\dot M=1.65\times 10^{11}M_{\rm \odot}/\, 10^6{\rm yr}.
\ee
For a fraction $f=0.01$ of the accretion mass to be converted to energy, the luminosity becomes for $\rho(\infty)=5\times 10^{-11}\,{\rm g/cm^3}$:
\be
L=f{\dot M}c^2=6.54\times !0^{49}\, {\rm erg/s}.
\ee
We see from (\ref{Eddlimit}) that this luminosity exceeds the Eddington limit luminosity $\sim 10^4L_{\rm Edd}$. Thus, we have to invoke a super-Eddington accretion rate to obtain a SMBH mass $M_{\rm BH}\sim 10^{10}M_{\odot}$ in the quasar lifetime $t_Q\sim 10^6$ yr~\citep{Haiman2004,Shapiro2005,VolonteriRees2006,Madau2014,Pezulli2016,Volonteri2015,Brightman2020}.

For a lower initial seed black hole mass $M_{\rm BH}=10^2M_{\odot}$, $\rho(\infty)=10^{-10}\,{\rm g/cm^3}$, $c_s(\infty)=10\,{\rm km/s}$ and $\beta=2.83$:
\be
\dot M=1.56\times 10^{29}\,{\rm g/s},
\ee
For a quasar lifetime $t_Q\sim 10^6$ yr:
\be
\dot M=2.62\times 10^9M_{\rm \odot}/\, 10^6{\rm yr},
\ee
and the luminosity for $f=0.01$ is
\be
L=f{\dot M}c^2=1.40\times !0^{48}\, {\rm erg/s}.
\ee
The super-Eddington accretion rate exceeds the Eddington luminosity rate, $L=100L_{\rm Edd}$.

For densities $\rho(\infty)\lesssim 10^{-13}$ the predicted luminosities are less than the Eddingon limit, $L_{\rm Edd}$. Moreover, the predicted growth of SMBHs to reach the detected masses requires super-Eddington accretion. However, the accrued mass during Bondi accretion cannot reach the detected mass $M_{\rm BH}\sim 10^9-10^{10}$ of SMBHs in the lifetimes of quasars and proto-galaxies. Measurement of the ambient gas and material density of the interiors of early universe quasars and proto-galaxies and the sizes of them can verify the SMBH evolution model. However, these measurements are difficult to make due to the distance of early universe quasars, proto-galaxies and SMBHs.

At late times in the universe with $z\sim 0$, the interstellar mass of an evolved galaxy will have a diluted gas density $\rho(\infty)=10^{-24}\,{\rm g/cm^3}$. For the speed of sound $c_s(\infty)=10\,{\rm km/s}$ and for the environment of the SMBH $M31^*$ with a mass $M_{\rm BH}=7\times 10^7M_{\odot}$ in the galaxy M31, we obtain with $\beta=2.83$ the accretion rate:
\be
\label{accreetgpersec2}
\dot M=7.63\times 10^{26}\, {\rm g/s}.
\ee
For a efficiency coefficient $f\sim 0.01$ the luminosity is
\be
\label{M81*}
L=f{\dot M}c^2=6.86\times 10^{45}\, {\rm erg/s},
\ee
that is within the Eddington limit luminosity:
\be
L_{\rm Edd}={\dot M}c^2=8.75\times 10^{45}\, {\rm erg/s}.
\ee

The SMBH in active galactic nuclei and quasars span a mass range, extending up to nearly $10^{11}M_{\odot}$~\citep{McConnell2011}. The heaviest black hole is associated with the quasar TON 618 with a mass $\sim 7\times 10^{10}M_{\odot}$~\citep{Shemmer2004}, while the second heaviest in the galaxy IC 1101 has a mass $\sim 4\times 10^{10}M_{\odot}$~\citep{Dullo2017}. This raises the question whether there is a mechanism that cuts-off larger black hole masses~\citep{Natarayan2009,Carr2020}. In our accretion scenario the accretion model depends on the density of ambient gas and material within a quasar or proto-galaxy and the available reservoir of this gas and material, allowing for a SPMBH to grow by Bondi accretion to a mass $M > 10^{11}M_{\rm \odot}$. For a SMBH with a mass $ M > 10^{12}M_{\odot}$ there would be a cut-off in the ambient density of a quasar or proto-galaxy of order $\rho > 10^{-10}\,{\rm g/cm^3}$. This would imply that the primordial cloud of gas that condenses to the primordial quasar and extra-galactic nucleus cannot exceed an upper limit on the gas density. Such an upper limit would be caused by the lack of strong enough gravity to collapse the primordial gas cloud, the complex thermal cooling of the gas as it condenses and on the stability of the primordial quasar and proto-galaxy.

\section{Conclusions}

To obtain SMBHs with a mass $M\sim 10^9-10^{10}$ at red shifts $z > 6$ ($\lesssim$ 1 Gyr after the Big Bang), one pathway is to invoke super-Eddington accretion through both disk accretion and Bondi accretion in the dense gas environment of quasars and proto-galaxies. Measurements of both black hole masses and bolometric luminosities for AGNs are difficult to achieve, so Eddington luminosity ratios are uncertain. However, the observed bolometric luminosity for most AGNs are estimated to be smaller than the Eddington value, $L_{\rm Edd}$~\citep{Kollmeier2006}. Super-Eddington accretion is believed to be observed in narrow-line Seyfert 1 galaxies, hosting SMBHs with accretion rates close to or greater than the Eddington limit~\citep{Pounds1995,Komossa2006,Jin2017}. Although local luminosities of quasars are below or close to the Eddington luminosity limit, we can consider it a strong possibility that in the early universe during the reionization era, when galaxies and quasars form, the explanation for the existence of SMBHs in young quasars with lifetimes only of order 100 thousand-1 million years is through the mechanism of super-Eddington accretion~\citep{Eilers2020}.

For accretion densities $\rho(\infty)\lesssim 10^{-13}\, {\rm g/cm^3}$, the Bondi accretion onto seed black holes with masses $M_{\rm BH}\sim 10^2-10^3M_{\odot}$ cannot reach SMBH masses $\sim 10^9- 10^{10}M_{\odot}$ within the ages of quasars and proto-galaxies in the early universe for $z > 6$, and for the predicted luminosities greater than or close to the Eddington luminosity limit, $L_{\rm Edd}$.

To produce SMBHs in the time of the age of young quasars with embedded black holes and an optimum ambient density in quasars, the luminosity $L$ must exceed the Eddington limit $L_{\rm Edd}$, and the rate of growth must exceed ${\dot M} > L_{\rm Edd}/fc^2$ for a period in the growth cycle. For the accretion rate in (\ref{accreetgpersec}), the super-Eddington accretion luminosity is
$L > L_{\rm Edd}$. Bondi accretion, together with Super-Eddington gas accretion for black holes with black hole seeds $M_{\rm BH}\sim 10^2-10^3$, is an attractive way to solve the rapid growth of SMBHs in young and longer life-time quasars and proto-galaxies for redshifts $z > 6$.

\section*{Acknowledgments}

I thank Martin Green and Viktor Toth for helpful discussions. This research was supported in part by Perimeter Institute for Theoretical Physics. Research at Perimeter Institute is supported by the Government of Canada through the Department of Innovation, Science and Economic Development Canada and by the Province of Ontario through the Ministry of Research, Innovation and Science.

\bibliography{refs}
\bibliographystyle{mn2e}

\end{document}